\documentclass[10pt,a4paper,twocolumn]{article}
\usepackage{hyperref}
\usepackage[margin=0.9in]{geometry}
\usepackage[version=3]{mhchem}
\usepackage{authblk}
\usepackage[version=3]{mhchem}
\usepackage[square,numbers,sort&compress]{natbib}
\usepackage{amsmath}
\usepackage{amssymb}
\usepackage{amsthm}
\usepackage{graphicx}
\usepackage{balance}
\usepackage{abstract}

\title{Reduced residence time of droplet impact on heated surfaces}

\author{Song Rong}
\author{Shiquan Shen}
\author{Tianyou Wang}
\author{Zhizhao Che\thanks{chezhizhao@tju.edu.cn}}
\affil{\small{State Key Laboratory of Engines, Tianjin University, Tianjin, 300072, China.}}

\date{}

\begin{document}

\twocolumn[
  \begin{@twocolumnfalse}
    \maketitle
    \begin{abstract}
The impact of droplets is a ubiquitous phenomenon, and reducing the residence time of the impact process is important for many potential applications. In this study of the impact dynamics on heated surfaces, we identify a mode of droplet bouncing (bouncing-with-spray mode) that can reduce the residence time significantly compared with the traditional retraction-bouncing mode. Comparing with other strategies to reduce the residence time, this approach induced by heat is simple and reliable. The reduction in the residence time is due to the burst of vapor bubbles in the liquid film, which results in the formation of holes in the liquid film and consequently the recoiling of the liquid film from the holes. A scaling law is proposed for the transition boundary between the retraction-bouncing mode and the bouncing-with-spray mode in the film boiling regime, and it agrees well with the experimental data. This model can also explain the transition between these two modes in the transition boiling regime.
    \end{abstract}
  \end{@twocolumnfalse}
]
\saythanks
\section{Introduction}\label{sec:sec1}
The phenomenon of droplet impact happens ubiquitously in nature and in a wide range of industrial applications which include but are not limited to spray cooling, painting, inkjet printing, and fuel-spray impingement in internal combustion engines \cite{Josserand2016ARFM, Yarin2005ARFMreview, Moreira2010howmuch}. For the impact of droplets on surfaces with different temperatures, the impact process could be significantly affected by the heat transfer or phase change \cite{Liang2017review, Gao2018}. Early studies of droplet impact on heated surfaces focused on the description of different impact morphologies in various impact regimes using high-speed imaging technique \cite{Chandra1991, Fujimoto2010, Staat2015} and the analysis of the ejected secondary droplets using phase Doppler anemometry (PDA) and image processing techniques \cite{Cossali2005, Moita2009, Moreira2007atomization}. When the droplet impacts on the heated surface, if the surface temperature is above the boiling point of the liquid, the droplet will go through a boiling process which is similar to that in pool boiling, including nucleate boiling, transition boiling, and film boiling \cite{Liang2016boiling}. When the surface temperature is high enough, a vapor layer will be generated underneath the droplet, i.e., the Leidenfrost effect \cite{Quere2013}, and the vapor layer acts as a thermal insulation layer between the droplet and the substrate. As a consequence, the vaporization rate of the droplet is reduced \cite{Bertola2015, Burton2012}. The Leidenfrost temperature is influenced by surface roughness \cite{Bernadin1996}, surface structure \cite{delCerro2012, Nair2014, Tran2013structured} and impact Weber number \cite{Celata2006}.

The contact time of a bouncing droplet is an important parameter in droplet impact process. For the axisymmetric impact of an inviscid droplet on a superhydrophobic surface, the contact time is bounded by the Rayleigh time scale \cite{Richard2002NatureScale}. Because the contact time controls the mass, momentum, and energy transfer between the droplet and the surface, it is important to reduce it in many applications, such as anti-icing, self-cleaning, corrosion-resistance, and maintaining surface dryness. Therefore, many efforts have been made to overcome the theoretical limit and reduce the contact time, for example, by using surfaces with ridges \cite{Bird2013NatRidge}, pancake bouncing on superhydrophobic surfaces patterned with lattices of submillimeter posts \cite{Liu2014Pancake}, asymmetric bouncing on curved surfaces \cite{Liu2015NatComCurved}, egg-shaped droplets \cite{Yun2018ResidenceTime}, etc.

In this study of the impact on heated surfaces, we call it \emph{residence time} instead of \emph{contact time} because the droplet only \emph{visually} contacts the substrate in the film boiling regime, and in fact, there is a vapor layer separating the droplet and the substrate. In nucleate boiling regime, because of the direct contact between the bottom liquid of the droplet and the substrate, the droplet can stick to the hot surface and evaporate, and the residence time of the droplet on the surface is strongly influenced by the surface temperature \cite{Breitenbach2017nucleateboiling}. Many theoretical models have been proposed to predict the contact time in the nucleate boiling regime \cite{Itaru1978, Tartarini1999, vanLimbeek2018}. In film boiling regime, a vapor layer forms between the droplet and the substrate, and it can act as a lubricant film. As a consequence, the droplet, if it has sufficient kinetic energy, may bounce from the substrate after the recoil, and therefore, the droplet residence time can be approximated as the period of a freely oscillating droplet \cite{YangGe2005, Kunihide1984, Tran2013structured, Ueda1979}. We find that a droplet bouncing mode, i.e.\ \emph{bouncing-with-spray} mode, caused by the burst of vapor bubbles can contribute to a dramatic reduction of the droplet residence time. Comparing with other strategies to reduce the residence time, this approach induced by heat is reliable by avoiding surface degradation and is easy to achieve by avoiding complex fabrication of surface microstructures or surface modification. Then a theoretical model is proposed for the transition between this mode and the traditional \emph{retraction-bouncing} mode. A scaling law is obtained and it agrees well with our experimental data.

\begin{figure}[tb]
  \centering
  % Requires \usepackage{graphicx}
  \includegraphics[width=0.9\columnwidth]{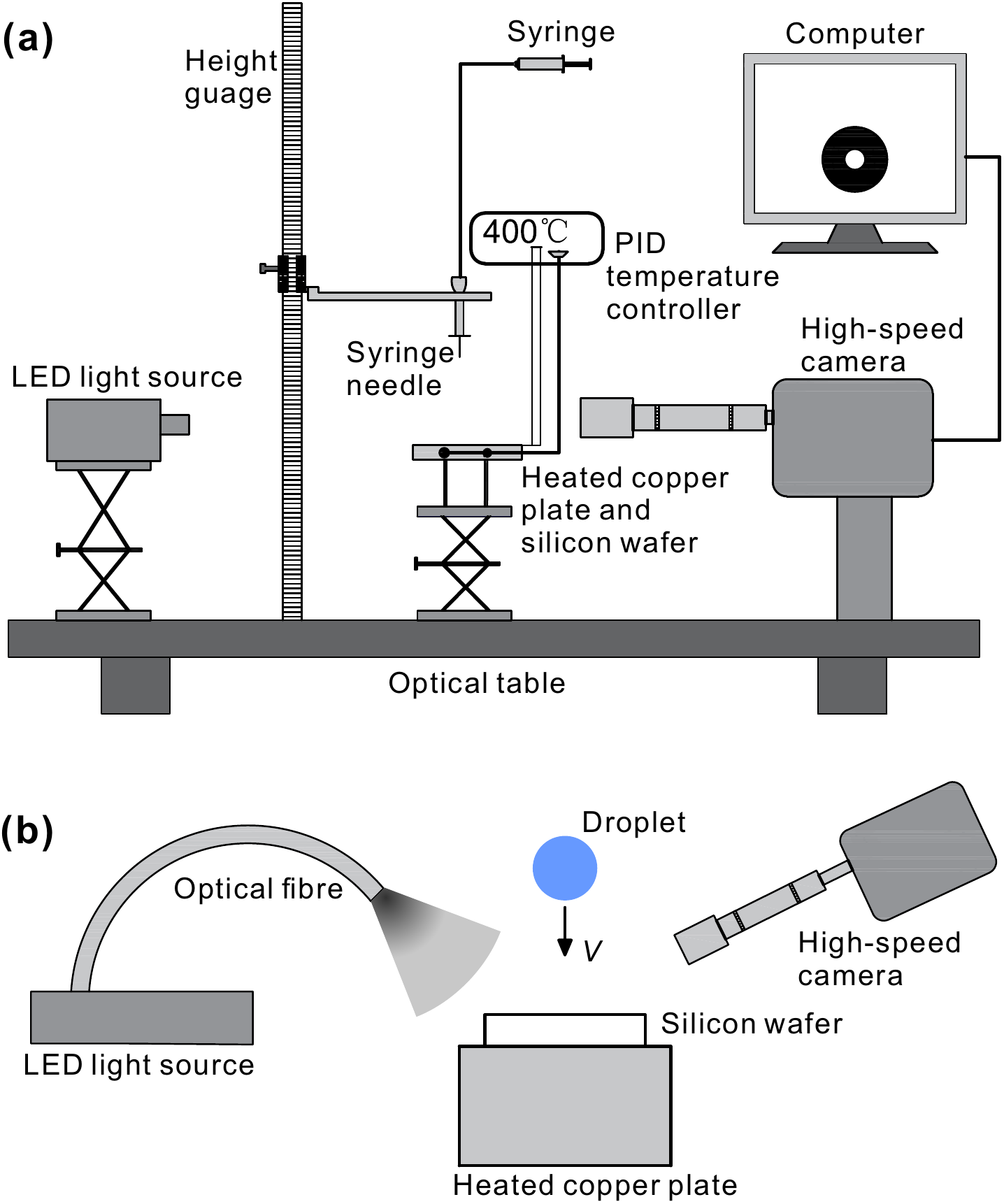}\\
  \caption{Schematic diagram of the experimental setup for droplet impact on a heated surface: (a) side-view, (b) aerial-view.}\label{fig:fig01}
\end{figure}

\section{Experimental method}\label{sec:sec2}
The experimental setup is schematically described in Figure \ref{fig:fig01}. Different droplet liquids (water, ethanol, and water/glycerol mixtures) were pushed through an injection tube at an extremely low speed (approximately 4 $\mu$l/s) by a syringe pump (Harvard Apparatus, Pump 11 elite Pico plus). Droplets formed at the tip of a blunt syringe needle and detached when the gravitational force exceeded the surface tension force. The droplet then impacted on a polished silicon substrate (silicon wafers, the roughness is less than 0.5 nm) which was heated by a copper holder. We placed two K-type thermocouples 1 cm beside the point of impact on the silicon surface to control the temperature with an accuracy of $\pm 1^\circ$C by a PID controller. We captured the side-view and aerial-view images of droplet impact evolution using a high-speed camera (Photron Fastcam SA1.1). The impact process was illuminated by an LED light source. The droplet size (typically $D_0=2.8$ mm for water/glycerol mixtures, 2.2 $\sim$ 3.4 mm for water, and 1.6 mm for ethanol), the impact speed ($V=0.9 \sim$  2.1 m/s), and the shape evolution of the droplet during the impact process, were measured from the high-speed images using a customized Matlab program. To change the viscosity of the impacting droplets, we used different liquids by changing the concentration of the glycerol in water/glycerol mixtures. The liquid properties are summarized in Table 1. The Weber number is used to quantify the ratio between the droplet's kinetic energy to its surface energy $W\!e\equiv {\rho {{D}_{0}}{{V}^{2}}}/{\sigma }$, where $\rho$ and $\sigma$ are the density and the surface tension, respectively. The $W\!e$ number was varied from 10 to 180 for different liquids. The Jakob number is used to indicate the ratio of the sensible heat to the latent heat of the liquid,
\begin{equation}\label{eq:eq1}
    Ja\equiv {{{c}_{p}}({{T}_{w}}-{{T}_{s}})}/{\ell },
\end{equation}
where $\ell$ is the specific latent heat of vaporization.

\begin{table*}[t]\label{tab:tab01}
\renewcommand{\arraystretch}{1.2}
\scriptsize
\centering
\caption{Properties of the liquids used in this study.}

    \begin{tabular}{lcccc}
\hline
Liquids                        & Density  & Surface tension  & Dynamic viscosity                        &Latent heat\\
                               & $\rho$ (kg/m$^3$)  & $\sigma$ (mN/m)  &$\mu$ (mPa$\cdot$s) 	       &$\ell$ (kJ/kg)\\
\hline
Water                          & 998                        & 72     & 1.00    &2257                           \\
{Water + glycerol (28 wt\%)}   & 1068                        & 70     &2.35     &                           \\
{Water + glycerol (34 wt\%)}   & 1083                        & 70     &3.00     &                   \\
{Water + glycerol (39 wt\%)}   & 1097                        & 70     &3.63     &                   \\
{Water + glycerol (56 wt\%)}   & 1143                        & 68     &9.00     &                   \\
{Water + glycerol (75 wt\%)}   & 1195                        & 66     &42.47    &                   \\
Ethanol                         &790                        &22     &1.07       &853                \\
\hline
    \end{tabular}
\end{table*}

\section{Results and discussion}\label{sec:sec3}
\subsection{Impact morphology and droplet residence time}\label{sec:sec3.1}
\begin{figure*}[tbh!]
  \centering
  % Requires \usepackage{graphicx}
  \includegraphics[width=1.5\columnwidth]{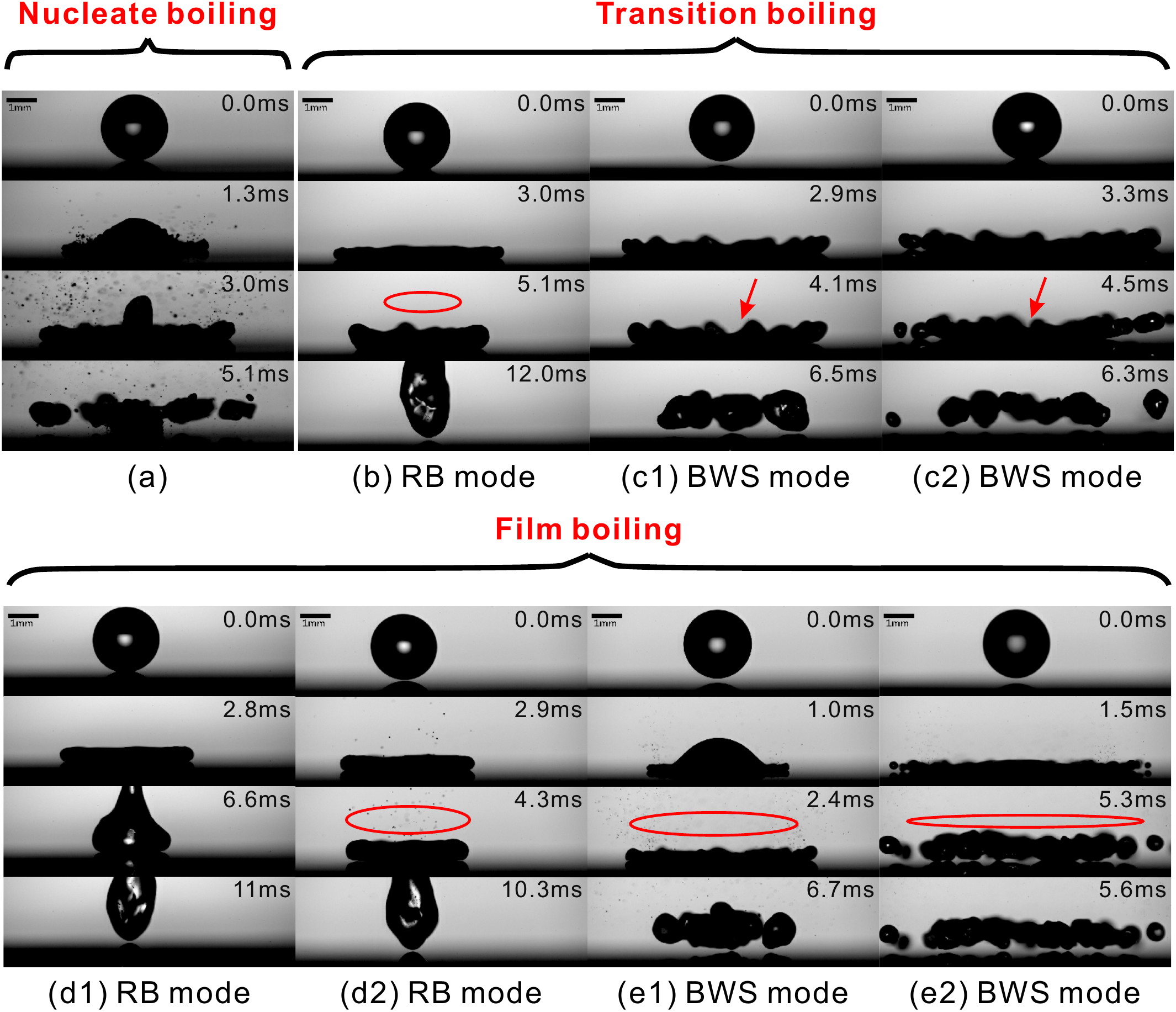}\\
  \caption{Morphologies of water droplet impacting on heated surfaces in three boiling regimes. The red circles and arrows highlight the spray-like ejection of tiny droplets. (a) Nucleate boiling, $W\!e=37$ and $T=260^\circ$C. (b) Retraction-bouncing mode in transition boiling regime, $W\!e=37$ and $T=380^\circ$C. (c) Bouncing-with-spray mode in transition boiling regime. $W\!e=69$ and $T=380^\circ$C for (c1), and $W\!e=105$ and $T=380^\circ$C for (c2). (d) Retraction-bouncing mode in film boiling regime. $W\!e=25$ and $T=400^\circ$C for (d1), and $W\!e=25$ and $T=440^\circ$C for (d2). (e) Bouncing-with-spray mode in film boiling regime. $W\!e=60$ and $T=440^\circ$C for (e1), and $W\!e=129$ and $T=440^\circ$C for (e2).}\label{fig:fig02}
\end{figure*}

Different impact morphologies were observed when varying the substrate temperature and the droplet speed and different kinds of liquids in three boiling regimes, namely nucleate boiling, transition boiling, and film boiling. Figure \ref{fig:fig02} shows the impact morphologies by taking water as a representative. These different morphologies result in a dramatic difference in the residence time of the droplet on the substrate. Here, the residence time is defined as the interval between the moments that the droplet visually contacts the substrate and that it visually detaches the substrate.

In the nucleate boiling regime, because of the direct contact between the bottom of the droplet and the heated surface, the liquid can boil so quickly that vapor bubbles form immediately upon impact, then rise through the droplet and break up at the free surface \cite{Tran2012}. As shown in Figure \ref{fig:fig02}a, the jet ejection due to the breakup of bubbles is followed by the inward surface wave propagation and the formation of tiny droplets created by the disintegration of the jets. After droplet spreading, the bottom liquid still sticks to the surface.

In the transition boiling regime, an unstable vapor film may exist between the droplet and the substrate so that the bottom of the droplet can contact the substrate locally. At a small impact speed (for example, $W\!e=37$ and $T=380^\circ$C, Figure \ref{fig:fig02}b), the droplet will recoil, then bounce off the substrate with spray-like ejection of tiny droplets (highlighted by the red circle in Figure \ref{fig:fig02}b). We name this mode of droplet bouncing in which droplet retracts its shape after the maximum spreading as \emph{retraction-bouncing} mode. In contrast, at a large impact speed (for example, $W\!e=69$ and $T=380^\circ $C, Figure \ref{fig:fig02}c1), the droplet can bounce off the substrate without obvious retraction before the departure of the droplet. This bouncing process is also accompanied by the spray-like ejection of tiny droplets (highlighted by the red arrow in Figure \ref{fig:fig02}c1). Therefore, we name this mode of droplet bouncing as \emph{bouncing-with-spray} mode. The residence time in this bouncing mode is much smaller than that in retraction-bouncing mode (6.5 ms vs.\ 12.0 ms). When the impact speed increases further (for example, $W\!e=105$ and $T=380^\circ $C, Figure \ref{fig:fig02}c2), we can see the formation and the breakup of liquid fingers at the rim of the droplet, indicating the splashing of the droplet due to the large droplet inertia \cite{Agbaglah2013, Allen1975, Roisman2006}. The central region of the droplet can still bounce off the substrate without obvious retraction. In addition, when comparing the residence time with that in the retraction-bouncing mode, it is also much smaller (6.3 ms vs.\ 12.0 ms). Therefore, we also categorize this mode of droplet bouncing as the bouncing-with-spray mode.

In the film boiling regime, stable vapor film will form underneath the droplet immediately upon the impact, i.e., the Leidenfrost effect \cite{Quere2013}, and the heat flux from the substrate to the droplet is reduced. In this regime, the retraction-bouncing and the bouncing-with-spray modes can also be seen. For the retraction-bouncing mode, Figure \ref{fig:fig02}d1 represents the case at a relatively low surface temperature compared with Figure \ref{fig:fig02}d2 (400$^\circ$C vs.\ 440$^\circ$C), and both $W\!e=25$. We can see that there is also ejection of tiny droplets (indicated by the red arrow in Figure \ref{fig:fig02}d2) at high surface temperatures and the distribution of the tiny droplets is more uniform than that in the transition boiling regime. It can be explained as that although there exists a stable vapor film, as the surface temperature increases, the growth rate of the vapor bubbles becomes higher so that it will be much easier for bubbles to rise through the droplet and burst on the free surface. On the other hand, with the impact speed increasing, the liquid film is thinner, and bubbles can rise to the free surface and burst as well \cite{Tran2012}. Then when the impact speed or the surface temperature increases further (for example, $W\!e=60$ and  $T=440^\circ$C, Figure \ref{fig:fig02}e1), the bouncing-with-spray mode can be seen in the film boiling regime. When $W\!e=129$ and $T=440^\circ$C, the splashing behavior and the bouncing-with-spray mode still occur as shown in Figure \ref{fig:fig02}e2.

\begin{figure*}[tb]
  \centering
  % Requires \usepackage{graphicx}
  \includegraphics[width=0.7\textwidth]{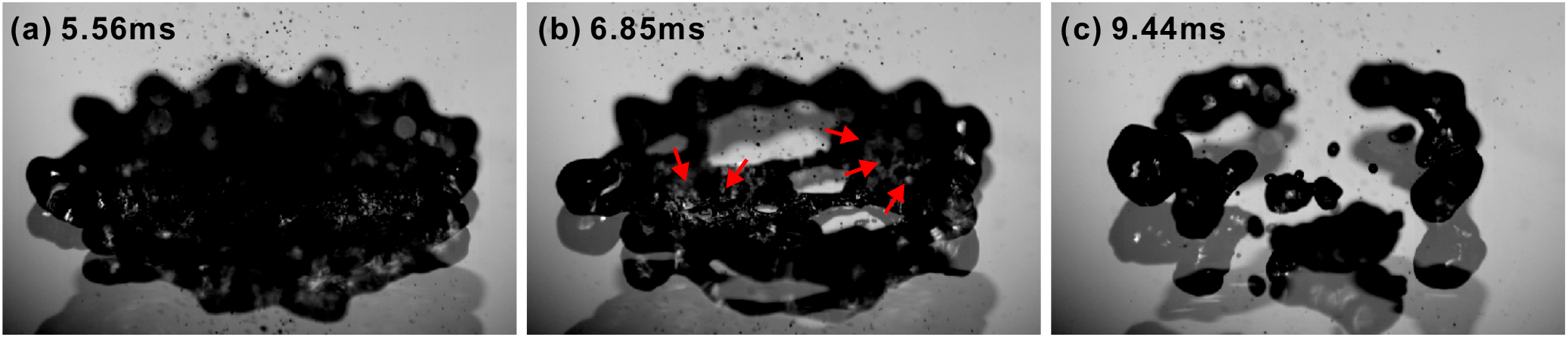}\\ %0.8, 0.45\textwidth
  \caption{Image series of the burst of vapor bubbles. $W\!e=82$ and $T=450^\circ $C.}\label{fig:fig03}
\end{figure*}

The residence time in the bouncing-with-spray mode is much smaller than that in the retraction-bouncing mode in both transition and film boiling regimes. To understand the mechanism of the reduction of the droplet residence time, Figure \ref{fig:fig03} shows a series of images of the bouncing-with-spray mode from an aerial view. During the impact process, vapor bubbles burst at random sites of the liquid film, producing holes one by one in the liquid film accompanied by several tiny droplets. The holes expand rapidly, and the liquid film recoils from these holes to several fragments (indicated by the red arrows) then bounces off the surface. The recoiling distance is reduced and so is the coiling time. Therefore, the residence time of the droplet is shortened significantly. This mechanism is similar to that of the reduction in the residence time during the bouncing of droplets on superhydrophobic surfaces with ridges \cite{Bird2013NatRidge}, where the ridges cause the fragmentation of the droplet and decrease the recoiling distance, and consequently reduce the residence time. In this study, the fragmentation of the droplet is induced by the burst of the vapor bubbles generated in the droplet fluid. Therefore, it is always accompanied by the spray of tiny droplets during bubble bursting. This mode of droplet bouncing and the consequent reduction in the residence time were observed for all kinds of liquids used in our experiments, and a detailed analysis is provided in the next section \ref{sec:sec3.2}.

\begin{figure}[tb]
  \centering
  % Requires \usepackage{graphicx}
  \includegraphics[width=\columnwidth]{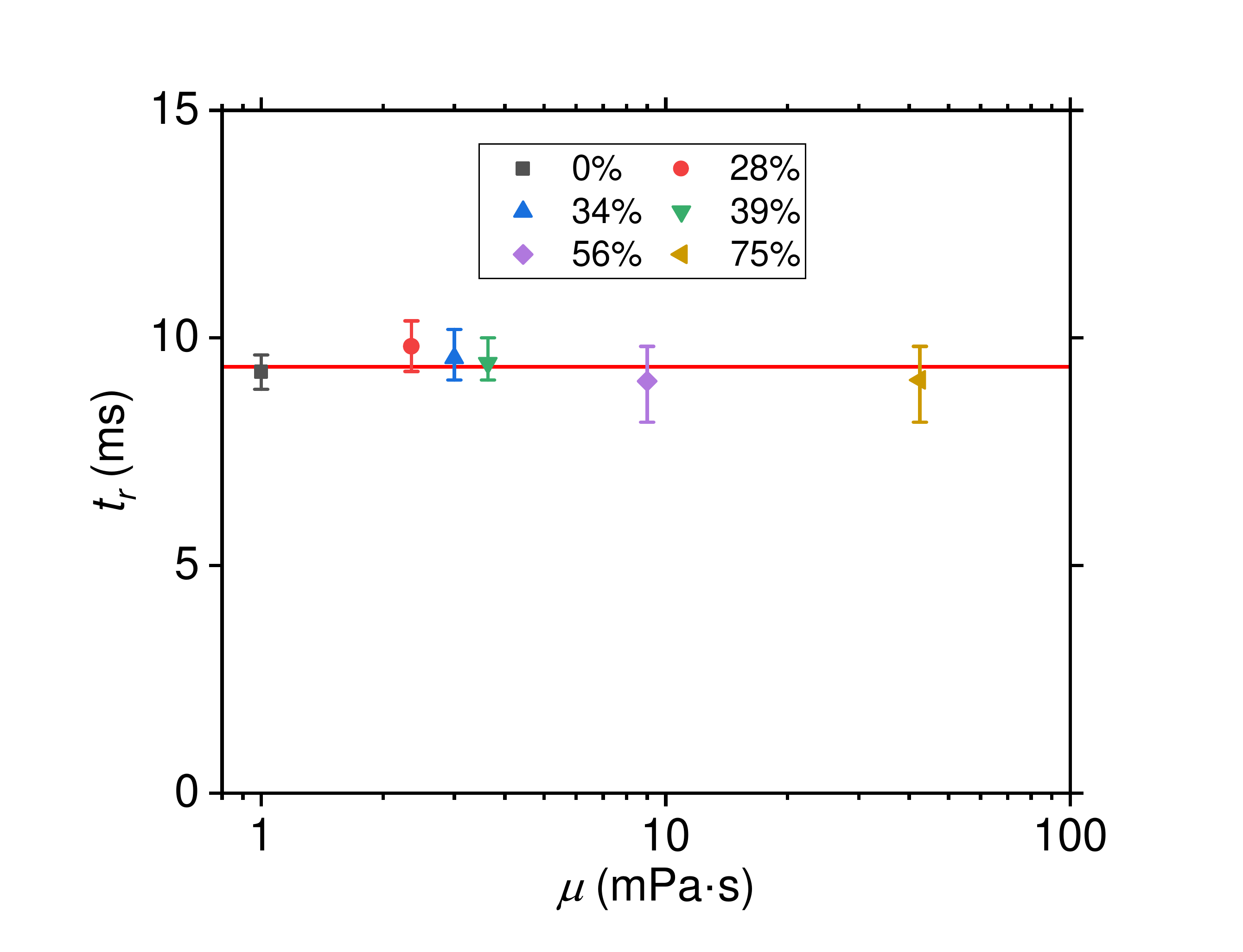}\\
  \caption{Effect of liquid viscosity on the residence time of the bouncing-with-spray mode. $D_0=2.8$ mm, $V=1.2$ m/s.}\label{fig:fig04}
\end{figure}

To determine the key factors that affect the bouncing-with-spray mode and its residence time, we varied the droplet viscosity by mixing glycerol with water at different concentrations. By using droplet fluids with different viscosities while fixing $V$ and ${{D}_{0}}$, we find that the residence time of the bouncing-with-spray mode is independent of the viscosity (see Figure \ref{fig:fig04}). This indicates that the viscous force does not play an important role in this process.

\begin{figure}[tb]
  \centering
  % Requires \usepackage{graphicx}
  \includegraphics[width=1.1\columnwidth]{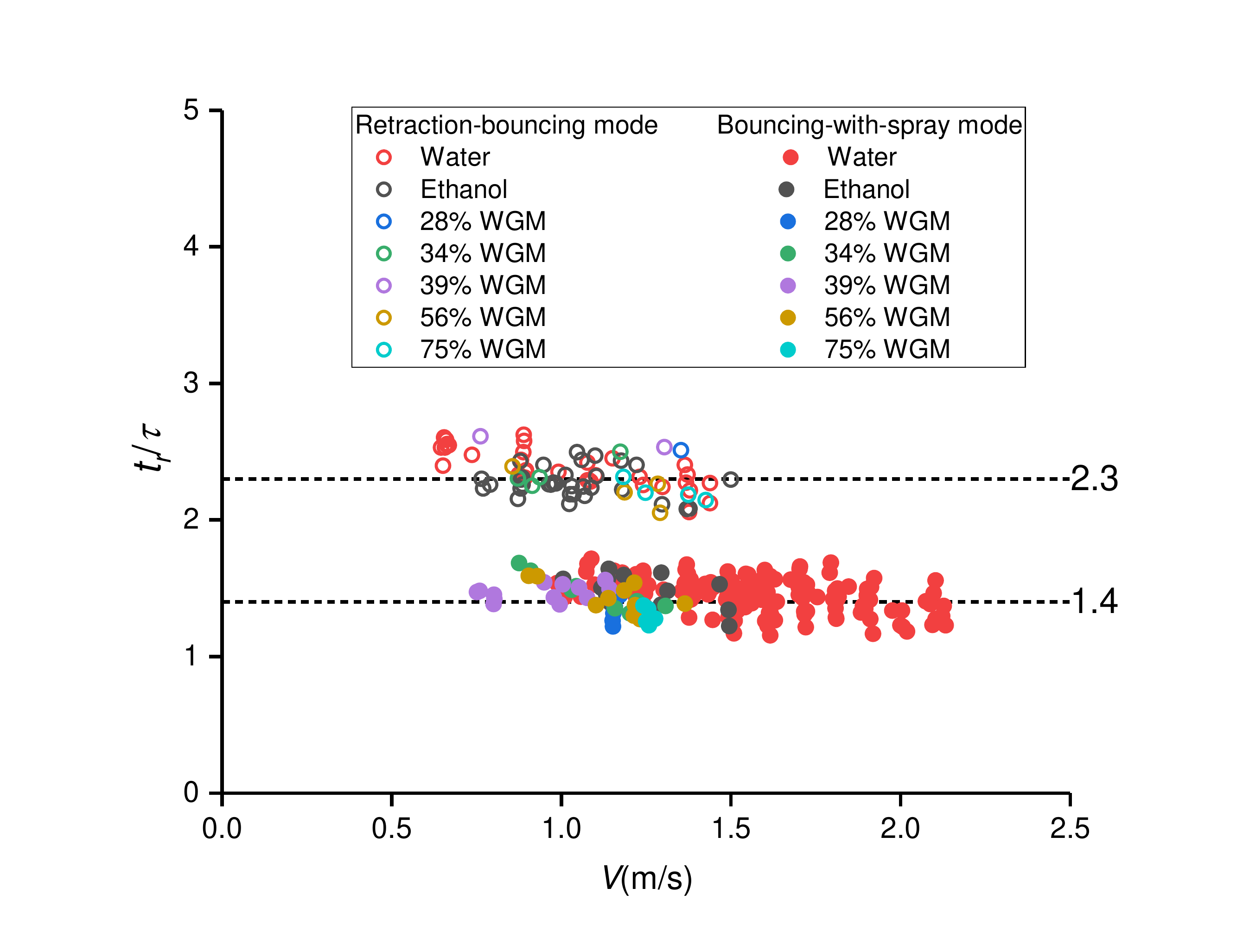}\\ %0.65
  \caption{Comparison of the dimensionless residence time in the retraction-bouncing mode (hollow circles) and in the bouncing-with-spray mode (solid dots). `WGM' means water-glycerol mixture. }\label{fig:fig05}
\end{figure}

For droplet impact on superhydrophobic surfaces or on superheated surface in the film boiling regime, the residence time in the retraction-bouncing mode is affected by the droplet radius, and scales with the inertial and capillary timescale:
\begin{equation}\label{eq:eq2}
  {{\tau}}=\sqrt{{\rho {{R}_{0}}^{3}}/{\sigma }}.
\end{equation}
By normalizing in this way, the residence time from past experiments are larger than the period of an oscillating droplet \cite{Rayleigh1879}, that is, ${\pi }/{\sqrt{2}}\approx 2.2$, and vary from 2.2 to 3.2 \cite{Richard2002NatureScale, Tran2013structured, Biance2006liquidShock, Chen2007dieseldrop}. In our experiments, the residence time is about 2.3 for retraction-bouncing mode, and to facilitate comparison on residence time reduction, the residence time in bouncing-with-spray mode is also expressed as the dimensionless form, which is about 1.4, as shown in Figure \ref{fig:fig05}, so the residence time reduction in our experiment is about 40\%.

\subsection{Residence time reduction mechanism and transition of bouncing modes}\label{sec:sec3.2}
\begin{figure*}[tbh!]
  \centering
  % Requires \usepackage{graphicx}
  \includegraphics[scale=0.55]{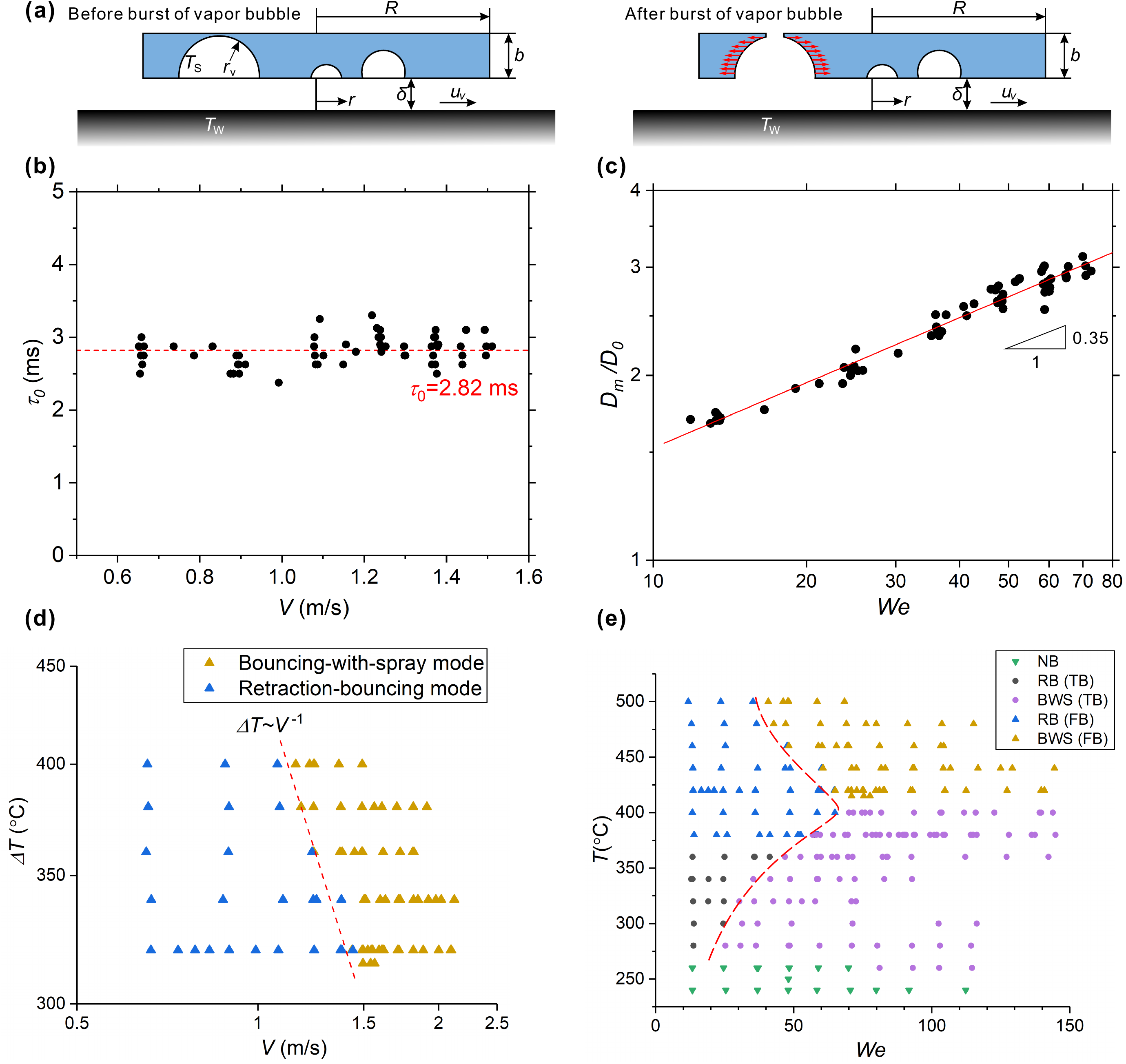}\\
  \caption{(a) Schematic diagram for the theoretical model for the burst of vapor bubbles. (b) Effect of impact speed on the time of maximum spreading diameter of water droplets, showing a constant ${{\tau }_{0}}$. (c) log-log plot for the normalized maximal spreading diameter of water droplets versus $W\!e$ in the film boiling regime, showing ${{{D}_{\max }}}/{{{D}_{0}}\sim{\ }{{W\!e}^{n}}}$. (d) $\Delta T\sim{\ }{{V}^{-1}}$ boundary line between the bouncing-with-spray mode and the retraction-bouncing mode in the film boiling regime according to Eq.\ (\ref{eq:eq17}) for water droplets at $D_0=2.2$ mm. (e) Regime map for a water droplet impacting on a heated silicon wafer and the transition between the retraction-bouncing mode and the bouncing-with-spray mode. The red dashed line is only for eye guidance, $D_0=2.2$ mm. `NB' means nucleate boiling regime, `TB' means transition boiling regime, `FB' means film boiling regime, `RB' means retraction-bouncing mode, and `BWS' means bouncing-with-spray mode.
  }\label{fig:fig06}
\end{figure*}

In the bouncing-with-spray mode, the vapor bubbles in the liquid film burst, resulting in holes in the liquid film and the spray of tiny droplets, as shown in Figure \ref{fig:fig03}. Then the liquid film recoils from these holes, and the reduction in the recoiling distance leads to the reduction in the recoiling time and finally contributes to the dramatic reduction in the droplet residence time. To further quantitatively understand the mechanism of the bouncing-with-spray mode and its effect on the reduction in the residence time, here we analyze the transition between the retraction-bouncing mode and the bouncing-with-spray mode in the film boiling regime. During the impact process, as the droplet expands, the size of the vapor bubble increases, and the film thickness decreases. The critical condition for the transition boundary between the retraction-bouncing mode and the bouncing-with-spray mode should correspond to a scenario that the radius of the vapor bubble is approximately equal to the minimum thickness of the film, which corresponds to the moment of the maximum droplet expansion, as shown in the schematic diagram in Figure \ref{fig:fig06}a. If the bubble radius is smaller than the minimum film thickness, the film will survive without hole formation. In contrast, if the bubble radius is larger than the film thickness, the bubble should have burst before the maximum droplet expansion, leading to the formation of holes in the film.

The formation and the growth of the vapor bubbles are due to the heat transfer from the substrate to the droplet fluid. Considering the conservation of energy, the amount of heat transfer from the substrate should be approximately equal to the vaporization heat of the bubble
\begin{equation}\label{eq:eq6}
  h\Delta T\Delta \tau A\sim \frac{2}{3}\pi r_{v}^{3}\ell {{\rho }_{v}},
\end{equation}
where $h$ is the average heat transfer coefficient, $\Delta T \equiv {{T}_{w}}-{{T}_{s}}$ is the degree of substrate superheat, $T_w$ and $T_s$ are the wall temperature and saturation temperature respectively, $A$ is the heat transfer area to the vapor bubble, $\ell$ is the specific latent heat of vaporization, and the subscript $v$ means `vapor'. The time interval for heat transfer $\Delta \tau$ can be approximated by the time for the droplet to reach its maximum lateral expansion ${{\tau }_{0}}$, which is independent of the impact speed (see Figure \ref{fig:fig06}b) and has been demonstrated to depend on the droplet oscillation period \cite{Biance2011fragmention}. Under the critical condition, the radius of the vapor bubble ${{r}_{v}}$ is approximately the minimum thickness of the liquid film ${{b}_{\min }}$, which has
\begin{equation}\label{eq:eq7}
  {{b}_{\min }}\sim{\ }{{Re}^{-{2}/{5}}}{{D}_{0}},
\end{equation}
as obtained in Refs.\ \cite{JensEggers2010, Alexander2005}.

Now we need to determine the average heat transfer coefficient $h$ in Eq.\ (\ref{eq:eq6}). We assume the vapor layer has a uniform thickness $\delta $. Because the heat transfer in the vapor layer is dominated by conduction as in other film boiling conditions \cite{Labeish1994}, the heat flux per unit surface area across the vapor layer is ${{{\lambda }_{v}}\Delta T}/{\delta }$, where $\lambda_v$ is the thermal conductivity of the vapor. Therefore a mass vaporization rate of droplet fluid is ${{{\lambda }_{v}}\Delta T}/{\left( \delta \ell  \right)}$. The mass conservation at a radial position $r$ of the vapor layer gives
\begin{equation}\label{eq:eq8}
  \pi {{r}^{2}}\frac{{{\lambda }_{v}}\Delta T}{\delta \ell }\text{=}2\pi ru\left( r \right)\delta {{\rho }_{v}}.
\end{equation}
Then the radial vapor velocity is
\begin{equation}\label{eq:eq9}
  u(r)=\frac{{{\lambda }_{v}}\Delta T}{2{{\rho }_{v}}\ell }\frac{r}{{{\delta }^{2}}}.
\end{equation}
Because of the small thickness of the vapor layer, the flow in the vapor layer can be regarded as Poiseuille flow. By considering the radially outward viscous flow of the vapor, the mean velocity across the vapor layer can be given as
\begin{equation}\label{eq:eq10}
  u(r)=-\frac{{{\delta }^{2}}}{12{{\mu }_{v}}}\frac{dp}{dr}.
\end{equation}
Then we can obtain the pressure distribution in the vapor layer by integration
\begin{equation}\label{eq:eq11}
  p(r)={{p}_{a}}+\frac{3{{\lambda }_{v}}{{\mu }_{v}}\Delta T}{{{\rho }_{v}}\ell }\frac{{{R}^{2}}-{{r}^{2}}}{{{\delta }^{4}}},
\end{equation}
where ${{p}_{a}}$ is atmospheric pressure, $R$ is the radius of the liquid disk. By integrating the pressure difference ($p(r)-{{p}_{a}}$), we can obtain the lift force exerted on the droplet disk by the vapor layer, and let it equal the difference between the weight of the droplet disk and the buoyancy force, $\pi {{R}^{2}}b({{\rho }_{l}}-{{\rho }_{v}})g$, as shown in Figure \ref{fig:fig06}a. Then the thickness of the vapor layer can be derived as
\begin{equation}\label{eq:eq12}
  \delta ={{\left[ \frac{3{{\lambda }_{v}}{{\mu }_{v}}\Delta T{{R}^{2}}}{2{{\rho }_{v}}({{\rho }_{l}}-{{\rho }_{v}})b\ell g} \right]}^{\frac{1}{4}}}.
\end{equation}
Therefore, the average heat transfer coefficient is
\begin{equation}\label{eq:eq13}
  h=\frac{{{\lambda }_{v}}}{\delta }={{\left[ \frac{2\lambda _{v}^{3}{{\rho }_{v}}({{\rho }_{l}}-{{\rho }_{v}})b\ell g}{3{{\mu }_{v}}\Delta T{{R}^{2}}} \right]}^{\frac{1}{4}}}.
\end{equation}
In the critical condition for the transition between the retraction-bouncing mode and the bouncing-with-spray mode, the radius of the liquid disk $R$ in Eq.\ (\ref{eq:eq13}) can be approximated by the maximum expansion diameter
\begin{equation}\label{eq:eq14}
  {{{R}_{\max }}}/{{{R}_{0}}}={{{D}_{\max }}}/{{{D}_{0}}\sim{{W\!e}^{n}}},
\end{equation}
as demonstrated in Figure \ref{fig:fig06}c, where $n=0.35$. In fact, many studies on spreading scale \cite{Antonimi2013, Biance2006liquidShock, Liang2016boiling, Tran2012, Tran2013structured} have been performed, and the scaling exponent $n$ can be different for different surface structures. The exponent in our experiments is close to that in the study of Tran et al. \cite{Tran2012}, where $n=0.39$.

By substituting Eq.\ (\ref{eq:eq13}) into Eq.\ (\ref{eq:eq6}), and using Eqs.\ (\ref{eq:eq7}) and (\ref{eq:eq14}) for $b$ and $R$ at the critical condition respectively, we can get
\begin{equation}\label{eq:eq15}
  {{\left( \frac{\lambda _{v}^{3}\left( {{\rho }_{l}}-{{\rho }_{v}} \right)g\Delta {{T}^{3}}}{{{\mu }_{v}}{{\ell }^{3}}\rho _{v}^{3}} \right)}^{\frac{1}{4}}}{{\tau }_{0}}A\sim {{W\!e}^{\frac{n}{2}}}R{{e}^{-1.1}}R_{0}^{\frac{13}{4}}.
\end{equation}
By introducing a modified Grashof number ${{Gr}^{*}}\equiv {g{{\rho }_{v}}({{\rho }_{l}}-{{\rho }_{v}})R_{0}^{3}}/{\mu _{v}^{2}}$, the Prandtl number of the vapor ${{Pr}_{v}}\equiv {{{c}_{pv}}{{\mu }_{v}}}/{{{\lambda }_{v}}}$, and a modified Fourier number ${{Fo}^{*}}\equiv {{{\lambda }_{v}}{{\tau }_{0}}A}/({{{\rho }_{v}}{{c}_{\rho v}}R_{0}^{4}})$, we have the following dimensionless scaling relationship
\begin{equation}\label{eq:eq16}
  {{\left( {{{c}_{pv}}}/{{{c}_{pl}}} \right)}^{\frac{3}{4}}}{{\left( {{Gr}^{*}}{{Pr}_{v}} \right)}^{\frac{1}{4}}}{{Fo}^{*}}{{Ja}^{\frac{3}{4}}}\ \text{ }\!\!\sim {{W\!e}^{\frac{n}{2}}}{{Re}^{-1.1}}.
\end{equation}
For different kinds of liquids, we suppose that the heat transfer area to the vapor bubble $A$ can be regarded as a constant. Therefore, ${{Gr}^{*}}$, ${{Pr}_{v}}$ and ${{Fo}^{*}}$ are mainly dependent on the physical properties of the liquids and the initial droplet size, then for the same liquid and the same initial droplet size, the physical properties of the liquids and the time of droplet maximum expansion ${{\tau }_{0}}$ can be regarded as constants. Using $n=0.35$ for this study, we can obtain
\begin{equation}\label{eq:eq17}
  \Delta T\sim {{V}^{-1}}.
\end{equation}

To check the validity of this scaling, the experimental data in the film boiling regime are plotted in logarithm scale in Figure \ref{fig:fig06}d, and they agree well with the scaling in Eq.\ (\ref{eq:eq17}). This theory can also explain why the transition temperature increases with $W\!e$ between the bouncing-with-spray mode and the retraction-bouncing mode in the transition boiling regime (see the regime map in Figure \ref{fig:fig06}e). In the transition boiling regime, the heat flux from the heated surface to the droplet decreases with increasing the substrate temperature. Therefore, for a certain time ${{\tau }_{0}}$, the maximum size of the vapor bubbles will decrease with the substrate temperature, and it requires a thinner liquid film (a larger $W\!e$) for the threshold of the burst of the vapor bubble ${{r}_{v}}\sim{{b}_{\min }}$. %, i.e., the radius of the bubble is approximately equal to the thickness of the liquid film.
Therefore, in the transition boiling regime, the critical $W\!e$ increases with the substrate temperature.

\section{Conclusions}\label{sec:sec4}
%\textbf{Conclusions}
In summary, the impact of droplets on heated surfaces is studied, and a droplet bouncing mode caused by the breakup of the liquid film, i.e., bouncing-with-spray mode, can contribute to a dramatic reduction of the droplet residence time compared to the traditional bouncing mode, i.e., retraction-bouncing mode. Comparing with other strategies to reduce the residence time, this approach induced by heat is reliable by avoiding surface degradation and is easy to achieve by avoiding complex fabrication of the surface microstructures or surface modification. The residence time of the bouncing-with-spray mode is independent of liquid's viscosity, and the reduced residence time compared with that of the retraction-bouncing mode is about 40\%.

The reduction in the residence time is due to the burst of vapor bubbles formed in the liquid film, resulting in holes formation and recoiling of the liquid film from the holes. The reduction in the recoiling distance leads to the reduction in the recoiling time. We have investigated the residence time reduction mechanism quantitatively by proposing a simplified theoretical model considering the energy balance and a critical condition of the bubble burst ${{r}_{v}}\sim{\ }{{b}_{\min }}$. According to this theoretical model, a transition boundary between the bouncing-with-spray mode and the retraction-bouncing mode in the film boiling regime is proposed. For the same liquid and the same initial droplet size, the transition between these two modes in the transition boiling regime is discussed. There are many open questions in this area of droplet impact on heated surfaces and its residence time. For example, detailed numerical simulations and experimental measurements of the bursting bubbles in the bouncing-with-spray mode will be helpful for a better understanding of the impact process. This study not only provides physical insight into the mechanism of the impact dynamics but also can be helpful in the optimization of this process in the relevant applications.

\section*{Acknowledgments}
This work is supported by the National Natural Science Foundation of China (Grant No.\ 51676137), the Natural Science Foundation of Tianjin City (Grant No.\ 16JCYBJC41100), and the National Science Fund for Distinguished Young Scholars (No.\ 51525603).
%\end{acknowledgments}
{\footnotesize
\bibliographystyle{model1-num-names}
\bibliography{HeatedDropletImpact}\balance}

\end{document}